\pdfoutput=1
\PassOptionsToPackage{hyphens}{url}
\documentclass[10pt,twocolumn,letterpaper]{article}

\usepackage[top=0.55in, bottom=0.55in, left=0.60in, right=0.60in, columnsep=0.22in]{geometry}
\usepackage{lmodern}
\usepackage{mathptmx}
\usepackage{graphicx}
\usepackage{xcolor}
\usepackage{booktabs}
\usepackage{array}
\usepackage{cite}
\usepackage{orcidlink}
\usepackage{url}
\usepackage{hyperref}
\usepackage{titlesec}
\usepackage{caption}
\usepackage{fancyhdr}

\hypersetup{
    colorlinks=true,
    linkcolor=blue!70!black,
    citecolor=blue!70!black,
    urlcolor=blue!70!black
}

\renewcommand{\thesection}{\Roman{section}}
\renewcommand{\thesubsection}{\Alph{subsection}}
\titleformat{\section}{\normalfont\fontsize{10}{12}\bfseries\scshape\centering}{\thesection.}{0.5em}{}
\titlespacing*{\section}{0pt}{1.0ex plus 0.2ex minus .2ex}{0.45ex plus .1ex}
\titleformat{\subsection}{\normalfont\fontsize{10}{12}\itshape}{\thesubsection.}{0.5em}{}
\titlespacing*{\subsection}{0pt}{0.8ex plus 0.2ex minus .2ex}{0.4ex plus .1ex}

\begin{document}

\twocolumn[{%
\begin{@twocolumnfalse}
\begin{center}
    {\fontsize{17.5}{21}\selectfont\bfseries Automated Pre-Silicon Verification of\\ DDR5 and LPDDR5/6 Memory Controllers\par}
    \vspace{0.6em}
    {\fontsize{11}{13}\selectfont
    \textbf{Manan Patel}\,\orcidlink{0009-0004-0290-6225} \textit{(Member, IEEE)} and
    \textbf{Anirban Majumder}\,\orcidlink{0009-0002-3490-1395} \textit{(Member, IEEE)}\par}
    \vspace{0.25em}
    {\fontsize{9.5}{11.5}\selectfont\itshape
    Google LLC, Mountain View, CA, USA \quad---\quad Google IT Services India Pvt.\ Ltd., Bengaluru, India\par}
    \vspace{0.7em}
\end{center}

\begin{quote}
\noindent\textbf{\textit{Abstract}---External memory interfaces (such as LPDDR5/4 and DDR5) are essential components of modern mobile, cloud, and enterprise computing systems. While memory manufacturers focus on physical DRAM die development, the vast majority of semiconductor firms design custom application ASICs that require a dedicated Memory Controller to interface with these standardized external memories. Consequently, pre-silicon design verification of the Memory Controller RTL---acting as the Device Under Test (DUT) against a third-party DRAM Verification IP (VIP)---is a ubiquitous and critical challenge across the global semiconductor industry. This article presents an automated, pre-silicon configuration and closed-loop initialization framework for Memory Controller verification. The proposed solution parses JEDEC timing and configuration parameters directly from the DRAM VIP's database files (such as Denali SOMA files) to configure the Memory Controller DUT's registers, while a custom Mode Register Register Abstraction Layer (MR-RAL) tracks volatile DRAM VIP states in real-time. Furthermore, a dynamic, re-compilation-free PHY initialization flow randomizes interface parameters directly in the testbench, executes on-the-fly configuration generation via system calls, and parses the resulting register write sequences at runtime. By integrating these automated methodologies into pre-silicon verification flows, engineers can eliminate setup overhead, prevent false protocol violations, and enable comprehensive randomized testing of complex PHY and Memory Controller configurations.}

\vspace{0.4em}
\end{quote}
\vspace{0.5em}
\hrule
\vspace{0.8em}
\end{@twocolumnfalse}
}]

\section{Introduction}
\noindent External DRAM is an essential component of all modern mobile, cloud, and edge computing systems. While memory-die manufacturers focus on the physics and chemistry of memory cell arrays, the broader semiconductor industry focuses on designing custom application ASICs and System-on-Chips (SoCs) to accelerate specific workloads. To utilize standardized external DRAM (such as LPDDR5/4 or DDR5), these custom ASICs must integrate a dedicated Memory Controller. The Universal Verification Methodology (UVM) has become the standardized method for SoC verification, heavily relying on the integration of Verification IP (VIP) to model these complex external dependencies~\cite{ref1}. Validating these interfaces is increasingly complex (as summarized across JEDEC generations in Table~\ref{tab:generations}); for instance, LPDDR6 PHYs operating at 14,400~Mbps on advanced process nodes present severe signal integrity challenges, including simultaneous switching noise and inter-symbol interference caused by the simultaneous flipping of dozens of data lines on a parallel bus~\cite{ref2}. While physical signal integrity phenomena are validated using physical-level or SPICE-level simulations, the pre-silicon digital verification bottleneck lies in validating the complex Memory Reference Code (MRC) training algorithms that the digital controller executes to actively compensate for these channel distortions. Therefore, verifying the digital Memory Controller RTL against external DRAM protocol standards is a critical, universal bottleneck for almost every semiconductor company, except for specialized memory manufacturers~\cite{ref3}. Achieving successful Memory Reference Code (MRC) training requires utilizing specific controller logic to initiate transactions on the LPDDR/DDR bus to tune sampling timings for all signals, scaling boot speeds from 4800~MT/s up to 14,400~MT/s~\cite{ref2}. In these pre-silicon verification environments, teams pair the Memory Controller Device Under Test (DUT) with a third-party memory VIP to simulate physical DRAM chips. This interface is heavily standardized, often relying on documents like the DDR PHY Interface (DFI) Specification version~5.1, which describes signals, timing parameters, and training sequences for the interface between DDR memory controllers and PHY layers~\cite{ref5}.

\begin{table*}[!t]
\centering
\caption{Generational Evolution of JEDEC DRAM Architecture and Pre-Silicon Verification Complexity (DDR4 to LPDDR6)}
\label{tab:generations}
\footnotesize
\begin{tabular}{>{\raggedright\arraybackslash}p{1.55in} >{\raggedright\arraybackslash}p{1.05in} >{\raggedright\arraybackslash}p{1.15in} >{\raggedright\arraybackslash}p{1.15in} >{\raggedright\arraybackslash}p{1.25in}}
\toprule
\textbf{Architectural / DV Metric} & \textbf{DDR4 (JESD79-4)} & \textbf{DDR5 (JESD79-5~\cite{ref6})} & \textbf{LPDDR5/5X (JESD209-5~\cite{ref7})} & \textbf{LPDDR6 (JESD209-6~\cite{ref2})} \\
\midrule
Peak Data Rate per Pin & 3,200 MT/s & 6,400 MT/s & 6,400--8,533 MT/s & 10,667--14,400 MT/s \\
Channel Organization & $1 \times 64$-bit & $2 \times 32$-bit sub-channels & 16-bit multi-channel & Dual 12-bit sub-channels (24-bit) \\
Mode Registers (MRs) per Rank & 7 (MR0--MR6) & 199 (MR0--MR255) & 56--63 (DVFS/FSP banked) & 89 (Extended training/RAS MRs) \\
Frequency-Scaled Timing Params & $\sim 100$ parameters & $\sim 160$ parameters & $\sim 150$ parameters & $\sim 165$ parameters \\
PHY Init / Training Registers & $< 250$ registers & $> 1{,}200$ registers & $> 1{,}500$ registers & $> 2{,}000$ registers \\
Classical Bring-Up per Speed-Bin & 0.5--1 day (manual) & 2--3 days (manual) & 5--6 days (manual) & 6+ days (manual) \\
\textbf{Proposed Automated Flow} & \textbf{$< 5$ minutes} & \textbf{$< 5$ minutes} & \textbf{$< 5$ minutes} & \textbf{$< 5$ minutes} \\
\bottomrule
\addlinespace[2pt]
\multicolumn{5}{l}{\scriptsize\textit{Note:} LPDDR5 bring-up time measured on the authors' project; other PHY register and bring-up figures are approximate, based on the authors' experience.}\\
\end{tabular}
\end{table*}

However, because JEDEC specifications (such as DDR5 JESD79-5~\cite{ref6}, LPDDR5 JESD209-5~\cite{ref7}, and LPDDR6 JESD209-6~\cite{ref2}) outline more than a hundred interdependent timing parameters that scale dynamically with clock frequency, synchronizing configurations between the Memory Controller DUT, the PHY, and the DRAM VIP remains a major industry challenge. Manually translating these dense, highly technical parameters across different simulation models introduces a massive risk of human error. A single parameter mismatch can trigger days of debugging overhead, masking actual design bugs behind false protocol violations. To address this, the industry requires a shift from static, manual configuration to automated, dynamic synchronization frameworks.

\section{Background and Current Memory Controller Verification Challenges}
In a classical pre-silicon verification flow, memory timing and configuration parameters must be manually extracted and programmed across multiple independent databases to align the Memory Controller DUT, PHY, and the DRAM VIP. This manual mapping is highly vulnerable to human error; a single timing parameter mismatch between the DRAM VIP simulation model~\cite{ref8} and the Memory Controller DUT's configuration registers can result in false protocol violations, consuming valuable debugging overhead.

Furthermore, high-speed memory PHY IPs require thousands of individual initialization and training parameters. In traditional verification flows, these parameters are often hardcoded directly into compiled SystemVerilog source files (such as packages or header files included via compiler directives) prior to compile-time. If a Design Verification (DV) engineer needs to modify a PHY setting---such as disabling a training stage or tweaking an impedance value---they must regenerate these source files and re-compile the entire simulation database. This compile-time dependency severely limits regression scaling and prevents the verification of dynamic, run-time parameter changes, as any parameter modification imposes a 100\% recompilation penalty, grinding iterative debugging to a halt.

\begin{table}[!t]
\centering
\caption{Comparison of Classical vs.\ Automated Memory Verification Flows}
\label{tab:comparison}
\scriptsize
\begin{tabular}{p{0.92in} p{1.05in} p{1.05in}}
\toprule
\textbf{Verification Metric} & \textbf{Classical Static Flow} & \textbf{Automated Closed-Loop Flow} \\
\midrule
Parameter Extraction & Manual mapping from datasheets & Automated script parsing \\
Re-compilation Requirement & Required for every parameter tweak & Eliminated (Run-time dynamic) \\
MR Synchronization & Manual database polling & Dynamic MR-RAL mirroring \\
Bring-up Overhead & Multiple days per speed-bin & $< 5$ minutes per speed-bin \\
\bottomrule
\end{tabular}
\end{table}

An additional critical challenge in memory controller verification involves the synchronization of DRAM Mode Register (MR) configurations. The MR configuration state must be seen identically by both the Memory Controller DUT RTL and the DRAM VIP. The traditional verification approach is highly cumbersome: engineers must manually retrieve MR-related configurations from UVM configuration databases or static files. This static, multi-step process is prone to synchronization lag and state mismatches, rendering the verification of dynamic training sequences highly error-prone. Table~\ref{tab:comparison} illustrates the stark operational differences between this classical methodology and an automated closed-loop approach.

\section{Automated SOMA-to-Register Parameter Synchronization}
To eliminate manual mapping overhead, we implemented an automated pipeline that extracts JEDEC timing and configuration parameters directly from the DRAM VIP's Cadence Denali SOMA (Specification of Memory Architecture) files and programs them into the Memory Controller DUT. The simulation behavior of a Cadence Denali VIP is largely dictated by these files; a \texttt{.denalirc} text file controls general simulation flow, while the SOMA file defines protocol standards and drivable values. Because SOMA is a Cadence-standard format used to parameterize verification models, it serves as an authoritative, machine-readable source of truth for all memory timing requirements, including complex parameters such as CAS latency, row precharge time ($t_{\mathrm{RP}}$), and RAS-to-CAS delay ($t_{\mathrm{RCD}}$)~\cite{ref6,ref7}.

During the pre-compilation phase (integrated into the build flow via \texttt{Makefile}), an automated script parses these DRAM VIP SOMA files. To ensure the tools can locate these essential parameters, the simulation environment must properly configure the \texttt{\$DENALI} environment variable to point to the correct VIP root directory. SOMA files contain certified timing specifications expressed in physical time units (picoseconds or nanoseconds) as well as clock cycles. Because the Memory Controller RTL operates on a digital clock domain, any parameter defined in physical time must be strictly converted into discrete clock cycles before it can be programmed into the DUT registers. Furthermore, the parsing script is designed to handle various speed-bins automatically, ensuring that when the simulation frequency changes, all dependent timing parameters are dynamically recalculated without requiring manual intervention.

For parameters defined in physical time units, the utility normalizes the values and translates them into clock cycles based on the active simulation frequency using JEDEC rounding conventions. This translation is mathematically expressed in (\ref{eq:cycles}), where the resulting cycle count is always rounded up to ensure timing constraints are never violated by being too short:
\begin{equation}
    N_{\mathrm{cycles}} = \lceil T_{\mathrm{param}} \times f_{\mathrm{CLK}} \rceil .
    \label{eq:cycles}
\end{equation}
Here, $N_{\mathrm{cycles}}$ represents the calculated cycle count, $T_{\mathrm{param}}$ is the JEDEC parameter in physical time, and $f_{\mathrm{CLK}}$ is the active clock frequency.

For parameters that are already defined in clock cycles, the script bypasses this translation and passes the values directly. The resolved clock cycles are then automatically mapped to their respective memory-mapped registers inside the Memory Controller DUT, generating a SystemVerilog configuration package that is subsequently compiled and elaborated. This ensures that the Memory Controller DUT, the physical layer, and the DRAM VIP operate in perfect clock-cycle synchronization, preventing false timing drift failures and completely automating the initial parameter programming phase. By automating this SOMA-to-register synchronization, verification engineers can seamlessly switch between different memory speed grades, such as moving from LPDDR5 4800~MT/s to 9600~MT/s in LPDDR5T, simply by pointing the environment to a new SOMA file. This eliminates the days of manual datasheet transcription previously required for each new configuration.

Additionally, most DRAM vendors maintain SOMA configurations for their custom requirements, especially in cases where they deviate from the baseline JEDEC specifications. The automated SOMA-to-register configuration flow allows the DV environment to seamlessly migrate to vendor-specific timings.

\section{Synchronized Mode Register (MR) Verification and Modeling}
To resolve the state-matching bottleneck across DRAM boundaries, we architected a unified Mode Register read/write API and a customized Mode Register Register Abstraction Layer (MR-RAL) model to align the Memory Controller DUT and the DRAM VIP. Volatile register fields undergo hardware-driven changes independently of bus transactions, rendering them invisible to standard UVM prediction logic and necessitating active monitoring to keep the register abstraction layer synchronized~\cite{ref10}. Because DRAM Mode Registers are inherently volatile during initialization and training sequences, standard frontdoor polling is vastly insufficient.

\begin{figure}[!t]
    \centering
    \includegraphics{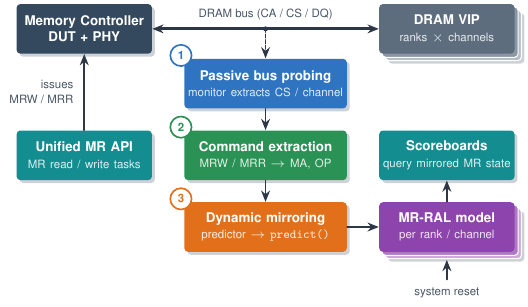}
    \caption{Automated Mode Register (MR) tracking and dynamic mirroring flow.}
    \label{fig:mr_tracking}
\end{figure}

The unified MR API exposes a set of tasks that programmatically handle and verify MR reads and writes. Instead of manually extracting configurations, this API manages transaction generation and automatically synchronizes the MR states between the Memory Controller DUT RTL and the DRAM VIP. To track these volatile DRAM states on-the-fly, a dedicated register group within the UVM RAL block is paired with a passive interface monitor. In UVM, the passive prediction mode updates the register model using transactions captured by the Universal Verification Component (UVC) monitor without connecting the model to the UVC sequencer~\cite{ref1}. This allows the testbench to silently observe the bus.

The synchronization process operates in three distinct phases, as depicted in Fig.~\ref{fig:mr_tracking}. First, through \textit{Passive Bus Probing}, the UVM monitor actively probes the physical interface bus to spy on transactions between the Memory Controller DUT and the DRAM VIP. To support complex multi-rank and multi-channel DDR5 topologies, the monitor also extracts active chip select (CS) and channel-routing signals. The passive predictor uses this routing metadata to map the extracted MR command to the corresponding rank-specific or channel-specific register block inside the hierarchical MR-RAL model. Second, during \textit{Command Extraction}, when the monitor detects a Mode Register Write (MRW) or Read (MRR) command, it extracts the target register address and the transferred data payload.

Finally, through \textit{Dynamic Mirroring}, explicit prediction is utilized. While auto-prediction updates mirror values during frontdoor access, explicit prediction uses a \texttt{uvm\_reg\_predictor} component connected to the monitor's analysis port to keep the register model updated~\cite{ref1}. In standard explicit prediction, the \texttt{uvm\_reg\_predictor} receives transactions from the monitor's analysis port, translates them using an adapter, and internally invokes the register's \texttt{predict()} method. Alternatively, for custom backdoor mirroring, the testbench can bypass standard prediction blocks by calling the \texttt{predict()} method directly:
\begin{center}
\scriptsize\texttt{mr\_ral\_block.MR\_reg[addr].predict(payload, .kind(UVM\_PREDICT\_DIRECT));}
\end{center}

This combination of a synchronized MR task API and dynamic MR-RAL mirroring ensures that check monitors and UVM scoreboards can query the exact state of the DRAM VIP at any moment. To maintain state alignment during hardware-driven asynchronous or system resets---which restore default register values without issuing bus transactions---the MR-RAL model is bound to the system reset signal. Upon reset assertion, the passive predictor triggers a reset of the RAL block, ensuring perfect state alignment is maintained across power cycles. Because these backdoor updates occur in zero simulation time, false out-of-sync checker flags are entirely eliminated, yielding a massive improvement in verification accuracy.

\section{Re-Compilation-Free PHY Initialization and Runtime Overrides}
To eliminate static pre-compiled configuration files and enable thorough randomized testing of the Memory Controller DUT, we developed an automated, closed-loop programmatic initialization flow. A standard SystemVerilog and UVM environment typically uses a Sequencer to generate transaction scenarios---such as testing boundary addresses and resets---which a Driver then maps to the DUT~\cite{ref1}. However, there may be cases where the DV testbench must be integrated with third-party IPs that use a separate compile or execution flow that is not inherently compatible with SystemVerilog/UVM. An example may be the usage of a third-party PHY IP, which requires a pre-compilation to generate static initialization sequences. Applying the standard UVM-based randomized testing architecture to thousands of static PHY initialization variables creates a massive performance bottleneck. Utilizing the standard UVM configuration database (\texttt{uvm\_config\_db} API) for run-time updates introduces significant simulation overhead because it requires costly string lookup operations for hundreds of elements~\cite{ref11}.

\begin{figure}[!t]
    \centering
    \includegraphics{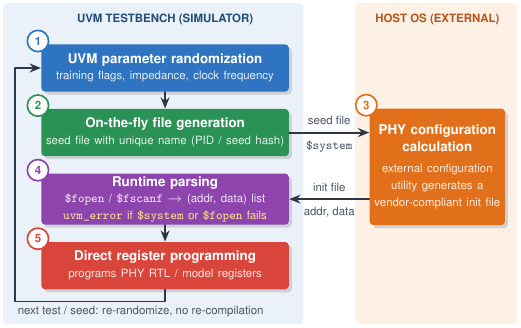}
    \caption{Closed-loop PHY initialization execution sequence.}
    \label{fig:phy_init}
\end{figure}

Therefore, the configuration process is driven directly from within the SystemVerilog/UVM testbench using a dynamic execution sequence that completely bypasses the standard UVM configuration database for PHY initialization. First, through \textit{UVM Parameter Randomization}, high-level PHY physical features (such as training flags, impedance settings, and target clock frequencies) are randomized using standard SystemVerilog architecture constraints.

Second, as illustrated in Fig.~\ref{fig:phy_init}, during \textit{On-the-Fly File Generation}, the testbench writes these randomized values to a temporary configuration seed and uses a SystemVerilog system call (\texttt{\$system}) to trigger an external configuration compiler utility. To ensure file isolation during concurrent regressions on a parallel compute cluster, the testbench appends the simulation process ID (PID) or a unique seed hash to the temporary file names, preventing race conditions and file collisions. Third, for \textit{PHY Configuration Calculation}, the utility runs on-the-fly to generate a uniquely isolated, vendor-compliant initialization text file. Fourth, a specialized SystemVerilog parsing function opens and reads this generated text file at runtime using standard file I/O operations (\texttt{\$fopen}, \texttt{\$fscanf}), extracting the hex register write sequence (addresses and data payloads). To guarantee execution robustness, the parsing logic evaluates the return status of the \texttt{\$system} call and validates the returned file descriptor from \texttt{\$fopen} before parsing, immediately throwing a UVM error if the external utility fails or the file is unreadable.

Finally, through \textit{Direct Register Programming}, the testbench applies the extracted sequence directly to program the active PHY RTL/model registers. This mechanism acts similarly to a pre-runtime signal forcing mechanism, allowing specific internal register states to be initialized before simulation begins, which synchronizes initial conditions across tests and bypasses standard protocol runtime overhead~\cite{ref11}. This closed-loop programmatic initialization bypasses static file generation entirely. Any PHY parameter or training threshold can be randomized and re-applied at run-time without requiring testbench re-compilation, representing a major efficiency improvement.

\section{Conclusion and Industry Implications}
Implementing this automated parameter synchronization and dynamic initialization framework yields substantial productivity gains in Memory Controller verification. In our deployment on an LPDDR5 memory controller, bringing up a new speed bin previously took a minimum of 5--6 days of manual timing-parameter and PHY configuration; with the automated SOMA-to-register and PHY initialization flows, it became plug-and-play once the initial parser setup and register-map configuration were in place. By extracting JEDEC parameters directly from authoritative SOMA files, semiconductor teams can significantly mitigate the human error associated with manual datasheet transcription, saving countless hours previously lost to debugging false protocol violations. While the current pipeline is designed around the proprietary Cadence Denali SOMA format, the underlying parameter synchronization methodology is generalizable. To adapt this framework for Verification IPs from other major vendors (such as Synopsys or Siemens), the front-end parser can be adjusted to read standardized XML schemas like IP-XACT~\cite{ref12}, leaving the core cycle-conversion and register-mapping logic intact.

By replacing pre-compiled static files with a dynamic system-call and runtime parsing flow, re-compilation bottlenecks are completely eliminated. This allows automated regressions to sweep hundreds of randomized physical configurations in a single run, rather than requiring engineers to manually re-compile the simulation environment for every minor impedance or timing tweak. The introduction of the MR-RAL passive tracking mechanism further guarantees that the verification environment maintains perfect state synchronization with the volatile DRAM VIP in zero simulation time.

Ultimately, these automated methodologies provide a highly scalable, robust blueprint that improves verification efficiency, enhances regression flexibility, and establishes a robust verification standard for high-bandwidth memory interfaces. As external memory standards like DDR5 and LPDDR5 continue to push the physical limits of signal integrity and timing margins, removing static verification bottlenecks will be paramount for delivering custom ASICs and SoCs to market on schedule.

\section*{Acknowledgment}
\noindent Generative AI tools were used in preparing this article. Gemini Next
(Google) and Claude Opus 5.5 (Anthropic) assisted with drafting and editing
text throughout the article, \LaTeX{} formatting, reference checks, and
counting the mode-register and timing-parameter entries in Table~I from the
cited JEDEC standards. Claude Opus 5.5 was used to create Figs.~1 and~2 as
vector graphics from the authors' descriptions and earlier draft figures. The
authors reviewed, edited, and verified all content and take full
responsibility for it.

\vspace{0.6em}
\noindent\rule{\columnwidth}{0.6pt}
\subsection*{Author Biographies}
\footnotesize

\noindent\textbf{Manan Patel} is a Senior ASIC Design Verification Engineer with Google LLC, Mountain View, CA 94043 USA. His research interests include high-speed memory subsystems, platform security verification, and automated verification methodologies. Patel has an MTech in VLSI design from Nirma University, Ahmedabad, India. He is a member of IEEE. Contact him at \href{mailto:mananpatel@google.com}{mananpatel@google.com}.

\vspace{0.4em}
\noindent\textbf{Anirban Majumder} is a Silicon Engineer with Google IT Services India Pvt.\ Ltd., Bengaluru, India. His research interests include DDR and LPDDR memory technologies. Majumder has a BTech and MTech dual degree in electrical engineering from the Indian Institute of Technology Kharagpur, Kharagpur, India. He is a member of IEEE. Contact him at \href{mailto:manirban@google.com}{manirban@google.com}.

\end{document}